# Ultrafast, green third-harmonic generation and strong-field phenomena in silicon-on-insulator nanoplasmonic waveguides

S. Sederberg and A. Y. Elezzabi

Ultrafast Optics and Nanophotonics Laboratory
Department of Electrical and Computer Engineering
University of Alberta
Edmonton, Alberta T6G 2V4 Canada

**Abstract**

The emergence of strong-field nanoplasmonics brings extreme laser field-matter interaction into the realm of nanoscale science, unveiling exciting new physics. Highly nonlinear interaction is enabled by tightly confined electric fields in nanoplasmonic structures, permitting use of optical fields from low-power laser oscillators. Here, we report the first demonstration of visible 517nm third harmonic generation in ultracompact nanoplasmonic waveguides on a silicon-on-insulator platform at an unprecedented conversion efficiency of $\sim 10^{-5}$. Exponential growth of broadband white light generation confirms a new strong-field phenomenon of ponderomotive force-driven electron avalanche multiplication. Using time-resolved experiments, we show that the strong nanoplasmonic field confinement allows nonlinear interaction to occur on an ultrafast timescale of $1.98 \pm 0.40$ ps, despite the long free-carrier lifetime in silicon. These findings uncover a new strong-field interaction that can be used in sensitive nanoplasmonic modulators and hybrid plasmonic-electronic transducers.

**Introduction**

Perhaps the most immediate and far-reaching application of nonlinear plasmonics is nanoplasmonic circuitry, where efficient excitation of nonlinearities is leveraged to modulate the propagating light signal[1-7]. Pioneering experimental investigations into nonlinear optical phenomena were performed in bulk materials, where transparent materials permitted a nonlinear interaction to accumulate under appropriate phase matching conditions, and lossy materials lent themselves to localized reflection interactions occurring within the skin depth of the constituent frequencies[8]. Subsequently, fibre optics and chip-scale waveguides enabled high electromagnetic intensities to be maintained over long interaction lengths, generating pronounced nonlinear signatures[9,10]. Amassing strong nonlinear signatures over short interaction lengths and at low input power has proven challenging, yet such devices are important elements of next-generation integrated optical nanocircuitry. Subwavelength confinement, electric field enhancement, and high sensitivity provided by nanoplasmonic structures make them excellent candidates for compact, all-optical circuitry with gate speeds exceeding those of conventional electronics by orders of magnitude. Furthermore, nanoplasmonic circuitry operating in the telecommunications band on a silicon-based platform would enable monolithic integration with electronic and silicon photonics technologies[11].

Indispensable in the electronics industry, silicon (Si) has accumulated a comprehensive infrastructure of growth and processing techniques that have enabled rapid advances in nanophotonic devices for routing, filtering, buffering, and modulating near-infrared electromagnetic radiation signals[12-14]. Third-order nonlinear effects occurring at the fundamental radiation frequency, $\chi^{(3)}(\omega)$, including two-photon absorption (TPA)[15,16], free-carrier absorption (FCA), Raman amplification,[17,18] the optical Kerr effect, four-wave mixing[19,20], cross-phase modulation[21,22], and self-phase modulation[23,24] have been studied extensively in silicon-on-insulator (SOI) waveguides. However, third-harmonic generation (THG) has remained elusive in SOI waveguides and has only been observed in a photonic crystal waveguide due to slow light-induced spatial pulse compression.[25] Third-harmonic generation has also been observed in plasmonic nanoparticles, but integration of these particles to optical circuitry has not yet been achieved.[26-28] Third-harmonic generation enhanced by strong confinement in a Si-loaded nanoplasmonic waveguide offers a novel platform for generating on-chip ultrafast green light sources.

Despite prior promising developments, the maximum integration density of Si photonic devices is limited by the diffraction limit, radiation loss that accrues around sharp bends, and cross-talk between adjacent waveguides. Moreover, the nonlinear interaction length of active devices is typically on the order of hundreds of microns, and such a footprint hinders its potential for large-scale integrated circuitry. The strong coupling of electromagnetic energy to the free electrons of noble metal features in nanoplasmonic waveguides enables highly confined waveguide modes at subwavelength dimensions and propagation around sharp bends. In order to maintain the possibility for monolithic integration with complimentary metal-oxide-semiconductor (CMOS) electronic devices and Si photonic devices, it is natural to consider Si-loaded plasmonic waveguides fabricated on an SOI platform. Electric field enhancement at the metal-Si interfaces would bolster $\chi^{(3)}$ nonlinear interactions, making possible the development of optical devices with advanced functionality.

Furthermore, Si-loaded nanoplasmonic waveguides enable advanced functionalities in integrated electronic-photonic circuitry, since the nanoplasmonic devices include high conductivity metallic layer(s). In this respect, hybrid plasmonic-electronic devices can be integrated within the current electronic device architecture where the nanoscale metallic-dielectric waveguides allow for the transport of both electrical and optical signals simultaneously on the same platform. This is of particular importance because the plasmonic device's size and the nonlinear interactions can be brought down to several nanometers, making them compatible with advanced complementary metal–oxide–semiconductor (CMOS) processor circuit architectures. The nanoplasmonic waveguides may also be efficiently interfaced to less lossy silicon-based plasmonic waveguides or silicon photonic waveguides for routing the signal over longer distances.

Silicon photonic crystal waveguides have been used to confine a large fraction of the input energy to the silicon waveguide core and to temporally compress pulses[25]. Silicon-loaded plasmonic waveguides provide an alternate means to achieve enhanced working intensities by confining the input energy to a smaller, sub-diffraction area. Although plasmonic losses limit the nonlinear interaction length to several microns, compensation provided by the electric field enhancement enables marked nonlinear interaction in a compact device.

Previously, optical excitation of free-carriers in the Si features of a surface plasmon grating coupler has been used to change the resonant coupling wavelength and modulate surface plasmon coupling to a planar Au film.[29] In fully-integrated Si-based nanoplasmonic waveguides, electric field enhancement at the metal-Si interfaces would bolster $\chi^{(3)}$ nonlinear interactions, yielding extreme nonlinearities that are typically only observable with complex chirped-pulse amplification systems. Nanoplasmonic antennas and tapered plasmonic waveguides may be used to enhance the intensity of laser radiation from a simple oscillator above the threshold ($\sim 10^{13} \text{Wcm}^{-2}$) required to observe high-harmonic generation as electrons are stripped from the atoms of noble gases, generating high-harmonics.[30]

Here, we report on several ultrafast nonlinear optical effects in Si-loaded plasmonic waveguides. We demonstrate THG at low average input powers in the range of 0.085-1.52mW ($\pm$0.03mW), a high power-normalized conversion efficiency of $(6.96 \pm 0.69) \times 10^{-10}$W$^{-2}$ (up to a maximum conversion efficiency of 2.29×10$^{-5}$) and a small device footprint of 0.43μm$^2$. The high field spatial gradient at the Si-Au interface produces a strong ponderomotive force that accelerates TPA-generated free-carriers, initiating avalanche electron multiplication. Broadband ($375 \leq \lambda \leq 650$nm) white light emission arising from the electron avalanche process grows exponentially with the input power. Pump-probe experiments reveal that the ponderomotive force sweeps free-carriers away from the probed region on a $1.98 \pm 0.40$ps timescale, demonstrating potential for hybrid integration of ultrafast nanoplasmonic devices with conventional CMOS electronics.

**Results**

Nanoplasmonic waveguide design

We investigate an SOI nanowire waveguide with a $t = 60$nm gold cap. A perspective view of the nanoplasmonic waveguide, with width, $w = 95$nm, height, $h = 340$nm, and length, $L = 5$μm, is shown in Fig. 1a. Figure 1b depicts the plasmonic mode profile, which has an effective index, $n_{eff} = 2.65$, and propagates a distance, $L_{prop} = 3.1$μm ($Loss = 1.4$dB/μm), before attenuating to $e^{-1}$ of its initial amplitude. The intensity distributions along the two longitudinal cross-sections in Figs. 1c-d portray the exponential decay of electromagnetic energy as the plasmonic mode propagates through the structure.

To measure ultrafast nonlinear optical dynamics, it is crucial to realize a means to excite a nonlinear interaction that is isolated to the structure of interest and not intertwined with a nonlinear excitation in an on-chip coupler. By eliminating the on-chip coupler altogether, free-space radiation may be coupled directly into the waveguide using a microscope objective and out-coupled using a lensed single-mode fibre (SMF). In order to realize this scheme, nanoplasmonic waveguides are fabricated on a narrow beam that is etched to a depth of $d_{etch}$ = 75μm, allowing for in-plane access with lensed SMFs. An artistic rendering of the sample along with the excitation and detection scheme is shown in Fig. 2a. A scanning electron micrograph (SEM) of a nanoplasmonic waveguide before integration onto the beam is shown in Fig. 2b. A beam including several nanoplasmonic waveguides embedded in silica cladding is shown in Fig. 2c, and the inset shows a cross-sectional SEM of the nanoplasmonic waveguide end facet. Cross-sectional dimensions of the Si core are measured to be $w \times h = 95 \times 340$nm and the Au cap has a thickness, $t = 60$nm.

Third-harmonic generation

Free-space radiation from a laser emitting $\tau_p \sim 84$fs pulses centered at $\lambda = 1550$nm is end-fire coupled into the waveguides using a 50× microscope objective. For peak input powers above $P_{peak}^{\omega} = 9.9$W, strong visible light emission is observed from the waveguides under normal room lighting. Scattered visible emission collected from above the sample by a 20× microscope objective and delivered to a charge-coupled device camera is shown in Fig. 2d. Bright white light emission with a strong blue component is visible at the input facet of the nanoplasmonic waveguide and distinct THG (green light) is observed only at the output facet. The light emission was present in every nanoplasmonic waveguide that was tested regardless of the waveguide length and no white light emission was observed from bare Si nanowires. Furthermore, a similar white light emission was observed from Si-loaded nanoplasmonic waveguides with an aluminium cap. Notably, some of these waveguides were longer than the propagation length of both the fundamental and the third-harmonic (TH) radiation. The $\lambda = 1550$nm laser pulses propagate through the nanoplasmonic waveguide with an effective wavelength, $\lambda_{eff,\omega} = 586$nm and decay with a characteristic length, $L_{prop} = 3.1$μm, while the TH radiation (500-530nm) propagates with $\lambda_{eff,3\omega} = 170$nm and decays over a length of 640nm ($1.1\lambda_{eff,\omega}$). The nonlinear interaction is essentially localized, as the TH radiation is absorbed as it interacts with the fundamental radiation on a scale of approximately one wavelength and the typical conditions of phase matching were not required to observe THG.

   The distinct difference in the spectral components of emitted light from the input and output facets of the waveguide (which are 4.5μm apart) provides clear evidence that we are not simply observing light emission from the incident free-space pulses exciting nonlinearities at the input facet of the waveguide and, in turn, scattering into the output fibre. Instead, the white light emission is stronger than the TH at the input facet. As the $\lambda = 1550$nm pulses propagate through the device, the nonlinear interaction evolves and as they reach the output facet of the waveguide, the TH emission is stronger than the white light emission, as indicated by the images and out-coupled spectra. This distinct evolution in the nonlinear interaction can only occur through a guided mode in the nonlinear Si core of the waveguide and would not be observed from scattered light from the input facet of the waveguide.

In order to investigate the nature of the visible light emission, the radiation emitted from the output facet of the nanoplasmonic waveguide is coupled to a lensed SMF, and delivered to a spectrometer. Full-wave simulations determine that ~4% of the visible light generated in the nanoplasmonic waveguide is coupled into the lensed optical fibre. The laser spectrum is shown together with the visible emission in Fig. 3a. The visible emission consists of a strong TH signal that spans $500\text{nm} \leq \lambda \leq 530\text{nm}$ and a broad white light background spanning $375\text{nm} \leq \lambda \leq 650\text{nm}$. The broad TH spectrum is further evidence that phase matching does not play a role in the THG.

A surface depicting the spectral power of the TH, $P_{out}^{3\omega}$, as a function of the average input laser power, $P_{in}^{\omega}$, is shown in Fig. 3b. The overlaid spectra depict a measured TH spectrum for $P_{peak}^{\omega} = 158 \pm 3\text{W}$, (dashed line in the figure), and a calculated TH spectrum using the free-space fundamental pulse spectrum. Regardless of $P_{in}^{\omega}$, the TH spectrum spans $500 \leq \lambda \leq 530\text{nm}$, and the peak wavelength is located at $\lambda = 517\text{nm}$. The total TH power is obtained by integrating the TH spectrum and a log-log plot of $P_{in}^{\omega}$ versus $P_{out}^{3\omega}$ is found to scale with a slope of $2.99 \pm 0.02$, as shown in Fig. 3c. Similarly, the slopes for specific wavelengths in the TH spectrum, $\lambda = \{512, 517, 520\}\text{nm}$, are found to be $3.04 \pm 0.01, 2.93 \pm 0.01$, and $3.16 \pm 0.04$, respectively. The $P_{out}^{3\omega} \propto (P_{in}^{\omega})^3$ dependence confirms that the green light originates from a $\chi^{(3)}(-3\omega;\omega,\omega,\omega)$ nonlinear optical interaction[31] and demonstrates that the entire TH spectrum grows uniformly with increasing input power. To date, there has been only a single observation of THG in a SOI device from a large (5μm-width×80μm-length) Si photonic crystal waveguide, where slow light propagation enhancement and tight modal confinement allowed for a THG conversion efficiency, $\eta^{THG} \approx 1\times10^{-7}$. Remarkably, the THG conversion efficiency from the nanoplasmonic structure is $\eta^{THG} = 2.29 \times 10^{-5}$, but in an ultracompact footprint (95nm×4.5μm) that is reduced by a factor of ~1400. For this nanoplasmonic structure, large electric field enhancement and tight spatial confinement are achieved at the Si-Au interface, which provide the necessary conditions for observing high efficiency THG.

White light emission and ponderomotive electron acceleration

It is important to elucidate the nature of the bright white light emission since its generation process is key in revealing the ultrafast carrier dynamics and their interaction with the highly-confined plasmonic field, $E_{SP}$. A surface plot showing the white light emission spectrum for varying $P_{in}^{\omega}$ is displayed in Fig. 3d. The overlay depicts the white light spectrum obtained for $P_{peak}^{\omega} = 165 \pm 3$W measured at the output facet of the nanoplasmonic waveguide, where the spectral contributions due to THG have been removed for clarity. The white light emission spans the range 375nm $\leq \lambda \leq$ 650nm (1.9eV $\leq h\nu \leq$ 3.3eV) and has a peak that lies within the THG spectrum (500nm $\leq \lambda \leq$ 530nm). Polarization measurements reveal that white light emission arises from coupling the TM-$E_{SP}$ mode into the nanoplasmonic waveguide. This and the absence of distinct atomic emission lines, increased scaling of the throughput light power with $P_{in}^{\omega}$, and lack of damage at the input facet, preclude the presence of polarization-independent surface breakdown.

As the femtosecond laser pulses propagate through the Si nanoplasmonic waveguides, electrons are excited from the valence band (VB) into the indirect bandgap ($E_g = 1.1$eV) at the conduction band (CB) X-valley via TPA. The average excess energy per electron is $E_{ex}$~0.5eV and the electron energy distribution lies near the bottom of the indirect CB edge. Since the energy of the white light emission spectrum (1.9eV $\leq h\nu \leq$ 3.3eV) is much higher than the two-photon energy, the excited electrons must gain additional energy and climb high into the CB directly after they are excited. However, electric fields acting upon the electrons supply additional energy and allow for radiative transitions at optical frequencies. Interestingly, a strong (>$10^9$V/m), *quasi-DC* electric field can be inherently provided by the highly spatially-inhomogeneous, oscillating plasmonic field. Once electrons are placed in the CB, they will interact with the strong evanescent (non-propagating, near-field) $E_{SP}$ field in the Si core of the nanoplasmonic waveguide. The nonlinear $E_{SP}$-electron interaction in the presence of a high $E_{SP}$ field gradient allows significant electron energy gain via ponderomotive (or cycle-by-cycle) acceleration. The basic principle of ponderomotive acceleration for an electron in an $E_{SP}$ field is illustrated in Fig. 4a. The high $E_{SP}$ propagating along the Au-Si interface is spatially inhomogeneous, rapidly decaying over a length scale of $\alpha_0 = 135$nm (sub-wavelength) in the Si core. As the $E_{SP}$ electric field oscillates over one cycle, it reverses its direction and a TPA excited electron in the CB experiences acceleration and deceleration during the positive and negative half cycles, respectively. If the $E_{SP}$ were spatially uniform, the electron would merely experience symmetric-but-opposing forces that would cancel over an optical cycle and no net kinetic energy would be gained. However, in a situation where the $E_{SP}$ field decays exponentially with an extreme spatial gradient, an electron with an initial kinetic energy, $E_{ex}$ that starts its motion at the Au-Si interface where the $E_{SP}$ field magnitude is highest, experiences asymmetric forces within the decay length (~135nm) of the $E_{SP}$ electric field. As the electron gains energy during the acceleration phase (positive cycle), it traverses into a region of a weaker $E_{SP}$ electric field before the electric field reverses during the negative cycle. As such, the decelerating force is weaker during the deceleration phase. Effectively, the electron is 'pushed' by the electric force in the direction of decreasing field amplitude by an amount that is larger than the subsequent cycle in which it is 'pulled' back. During successive rapid oscillations of the $E_{SP}$ field, the electron is pushed away from the interface in a cycle-by-cycle fashion, and the electron is imparted with a net velocity, $v(t)$, along the direction of decreasing electric field amplitude as illustrated in Fig 4a. With this pondermotive acceleration-deceleration being repeated over several oscillation cycles, a low-frequency (quasi-DC) ponderomotive force, $F_{pond}$, arises in a direction normal to the Au-Si interface and imparts a net kinetic energy gain to the electron.

Nanoplasmonic field-enhanced impact ionization

The transfer of kinetic energy among electrons occurs via impact ionization. This is a three-particle process that may be observed in Si when initial free-carriers (generated by TPA) are subjected to a high electric field. As the free-carriers gain energy from the ponderomotive potential, they undergo scattering interactions with bound electrons in the valence band of the semiconductor. A threshold energy of $E_t = 2.3\text{eV}$ must be imparted to the valence band electrons for impact ionization to take place[32]. It is important to note that the electrons may gain this energy before they undergo their first collision (~28nm) and their trajectory may be considered as ballistic during this time. Sufficiently energetic interactions promote new electrons to the conduction band, resulting in a new electron-hole pair. Secondary electron-hole pairs can attain a high energy in a similar manner and go on to multiply in an "avalanche mode", yielding exponential growth in the number of electrons. Electrons that undergo collisions before reaching $E_t$ may emit a photon depending on the nature of the collision.

A signature of the avalanche process would be exponential growth of the electron number as a function of the intensity of the plasmonic field or equivalently, an exponential growth of light emission as a function of $P_{in}^{\omega}$. As evidenced in Fig. 4b, the total white light power, $P_{wh}$, grows exponentially with $P_{in}^{\omega}$ according to the relation $P_{wh} \propto exp\ (\gamma P_{in}^{\omega})$, where $\gamma$ is the power growth coefficient (~2.38 ± 0.15mW$^{-1}$). The exponential power-dependence on $P_{in}^{\omega}$ is observed at all wavelengths within the spectrum, and the scaling trends for $\lambda = \{472, 537, 578\}$nm are shown in Fig. 4b. The distinct exponential growth signifies that the impact ionization mechanism has taken place due to ponderomotive electron acceleration. Notably, white light luminescence from the gold film would show a fourth-order dependency on $P_{in}^{\omega}$.

Twenty-one simulated electron trajectories in response to the $E_{SP}$ field are shown in Fig. 4c. Each of these electrons is introduced to the simulation at a different point in time, and with a different starting position and velocity direction (the velocity amplitude was kept constant at 0.5eV). Approximately 1.4×10$^5$ electron trajectories are simulated for three $E_{SP}$ strengths, $E_{SP} = \{1\times10^9, 3\times10^9, 5\times10^9\}$V/m and plots of the electron's excess energy gained from the ponderomotive potential are calculated, as shown in Fig. 4d. It is determined that a minimum field of $E_{SP}$~2.5×10$^9$V/m is required to accelerate electrons to an energy of $E_t = 2.3$eV. This is in excellent agreement with the estimated experimental electric fields in the structure, which are in the range 9.10×10$^8 \leq |E_{SP}| \leq$ 4.77×10$^9$ (±2.7×10$^8$)V/m. For $E_{SP}$ = 5×10$^9$V/m, the electron energy distribution reaches a maximum value of ~ 6eV. It should be noted that these plots do not represent the thermal energy distribution (Maxwellian distribution) of the free electrons, but rather the excess kinetic energy that the electrons gain from the plasmonic field via ponderomotive accleration. Further, there is not a simple direct relationship between the calculated electron energy spectrum and the observed white light spectrum. The electron energy spectrum is similar to the ones reported previously for photoelectron emission from metal films[33-36].

White light emission model

The emitted white light spectrum shown in Fig. 3d is featureless and resembles that observed from a Si diode, Si-junction gate field-effect transistor (JFET), and Si-bipolar junctions under high reverse bias voltage[37-39]. To describe our observations theoretically, we calculate the emission spectrum using a multimechanism model for photon generation by Si junctions in avalanche breakdown[40]. The three dominant emission mechanisms used in this model are depicted schematically in Fig. 5a: (i) direct interband (D-I) recombination between hot electrons and holes near $k = 0$; (ii) intraband transitions (I-T), or Bremsstrahlung emission; and (iii) indirect interband (I-I) recombination of electrons and holes. Figure 5b illustrates the multimechanism model applied to the experimental data. Clearly, the excellent agreement between the model and experiment confirms the physical origin of the light emission. The emission intensity at energies below $E = 1.9$eV is relatively weak, demonstrating that the majority of electrons are accelerated to high energies between collisions. Furthermore, the emission intensity drops off at energies greater than $E = 3.3$eV. Electrons that are accelerated to energies greater than $E_t$ (which have total energy of $E_t + E_g \sim 3.4$eV) and undergo a collision will initiate impact ionization and return to the conduction band edge. The absence of radiation at energies greater than 3.3eV provides direct evidence of the prominent impact ionization process.

Ultrafast nonlinear dynamics

Additional evidence signifying new underlying physics of the phenomenon of ponderomotive plasmonic field induced electron avalanche growth can be revealed by the time dependence of the electron dynamics. In the present nanoplasmonic waveguide, electrons are generated in the Si region via TPA and the majority of these electrons exist close to the Au-Si interface where the electric field is the most intense. In the presence of a strong $F_{pond}$, the electrons in the Si are accelerated and swept away very quickly in the $E_{SP}$ field ($\sim 10^9$ V/m), reaching a saturation velocity, $v_s \approx 1.3 \times 10^5$ m/s. Thus, within ~1.0ps, these electrons traverse a distance, $\alpha_0 = 135$nm from the Au-Si interface, and within ~2.6ps, they would cross the entire Si core ($h = 340$nm). Therefore, it can be expected that the measured electron dynamics take place on a timescale between these two values. Electron sweeping dynamics can be captured in real time via ultrafast time-resolved spectroscopy.

Ultrafast pump-probe experiments are performed on the nanoplasmonic waveguides in order to confirm the presence of TPA and FCA, and visualize the electron sweeping dynamics. Two experimental configurations are investigated: one where the polarization of the pump is parallel to that of the probe (co-polarized) and the other where their polarization is perpendicular (cross-polarized). An artistic rendering of the cross-polarized pump-probe characterization setup is shown in Fig. 6a. A pump-probe time domain trace obtained in the co-polarized configuration is shown in Fig. 6b. Interference fringes appear in the traces as the pulses begin to overlap and as the pulse peaks overlap, nonlinear loss due to TPA ($Loss \propto (P_{in}^{\omega})^2$) and FCA ($Loss \propto NP_{in}^{\omega}$, where $N$ = free-carrier density) produces a sharp dip in the transmission. Due to the quadratic power scaling, TPA dominates the interaction and the full-width at half maximum of the dip is dependent primarily on the pulse duration and is found to be ~367fs. Since both the pump and probe pulses are detected, this trace is essentially an autocorrelation of the laser pulse. Cross-polarized pump-probe experiments enable the pump radiation to be filtered out with a polarizer, providing more details of the nonlinear interaction and time-dynamics. First, the nonlinear scaling of the loss associated with TPA, FCA, and the $F_{pond}$ driven electron avalanche is investigated. A log-log plot of the total nonlinear loss versus $P_{in}^{\omega}$ is shown in Fig. 6c and is found to be linear with a slope of $2.52 \pm 0.12$. Time-domain traces for pulses propagated through a nanoplasmonic waveguide for varying pump power $P_{in}^{\omega} = \{142, 170, 198\} \pm 28\mu W$, are shown in Fig. 6d. Interference fringes appear in the traces as the pulses begin to overlap. As the pulse peaks coincide, a combination of TPA and FCA produce a sharp drop in the transmission. Focusing achieved by the *NA* = 0.85 microscope objective produces mixed polarization components at the waveguide input facet. Therefore, some components of the pump are aligned parallel to the probe, producing the interference fringes and providing ponderomotive acceleration. As $N$ decreases due to field sweeping and recombination (two-body, surface, Auger), the transmission signal recovers to its original amplitude. As shown in Fig. 6d, this recovery occurs on two timescales: $\tau_1 = 1.98 \pm 0.40$ps and $\tau_2 = 17.9 \pm 6.8$ps, which are much faster than the recovery time in a bare Si waveguide with a comparable size ($w \times h = 340 \times 340$nm), shown in the bottom of Fig 6d. The recovery tail of the bare Si waveguide makes a negligible recovery in the temporal window shown and longer scans reveal that the bare Si waveguide recovers on a timescale of approximately $\tau = 265$ps. The shorter lifetime is in excellent agreement with the expected $F_{pond}$ sweeping time ($\sim 1.0$ps $- 2.6$ps) required for electrons to travel $\sim 135$nm $- 340$nm away from the Au-Si interface to a region where they no longer interact with the probe pulse.

The longer time scale, $\tau_2$, is attributed to the free-carrier lifetime in Si. Reduction of the free-carrier lifetime to $\tau_2 = 17.9 \pm 6.8$ps in SOI nanowires requires advanced processing techniques such as ion-implantation or carrier sweeping with an integrated p-n junction[41,42]. In the present case, diffusion of the Au cap into the Si waveguide core leads to an increase the density of recombination centers and surface traps near the Si-Au interface, increasing the surface recombination velocity. Notably, military specification electronic circuits use Au interconnects, and it is anticipated that integration of Au features into CMOS electronics would not affect electronic performance. Alternatively, a non-diffusive plasmonic material, such as TiN, could replace the Au features.

**Discussion**

In this article, we report the first realization of THG and strong-field phenomena in a CMOS-compatible nanoplasmonic platform. We demonstrate plasmonically enhanced THG with a high conversion efficiency of $\sim 10^{-5}$. Electrons excited via TPA are accelerated to energies up to several eV by the ponderomotive potential that exists in the highly confined plasmonic field. Subsequent collisions with valence electrons drive impact ionization, which multiplies in an avalanche. Electrons that do not reach the threshold for impact ionization emit photons during collisions, and white light emission is observed to scale exponentially with the input power, confirming the exponential growth of avalanching electrons. Electron sweeping due to ponderomotive electron acceleration is visualized using pump-probe time domain spectroscopy and the sweeping time is measured to be $\tau_1 = 1.98 \pm 0.40$ps, confirming that the electrons reach saturation velocity. Pondermotive electron acceleration has been used as a scheme for the generation of ultrashort energetic free electrons bunches from metal films into vacuum[39-42], but has not been observed in semiconductors. This device operates at telecommunications wavelengths, occupies an ultracompact footprint of $0.43\mu m^2$ and is fabricated on a SOI platform, making it compatible with existing technologies. The high sensitivity of the electron avalanche process will enable the development of compact, sensitive optical circuitry and interfacing between plasmonic and electronic components. These finding offer a means to harness the potential of the emerging field of nonlinear nanoplasmonics.

**Figures**

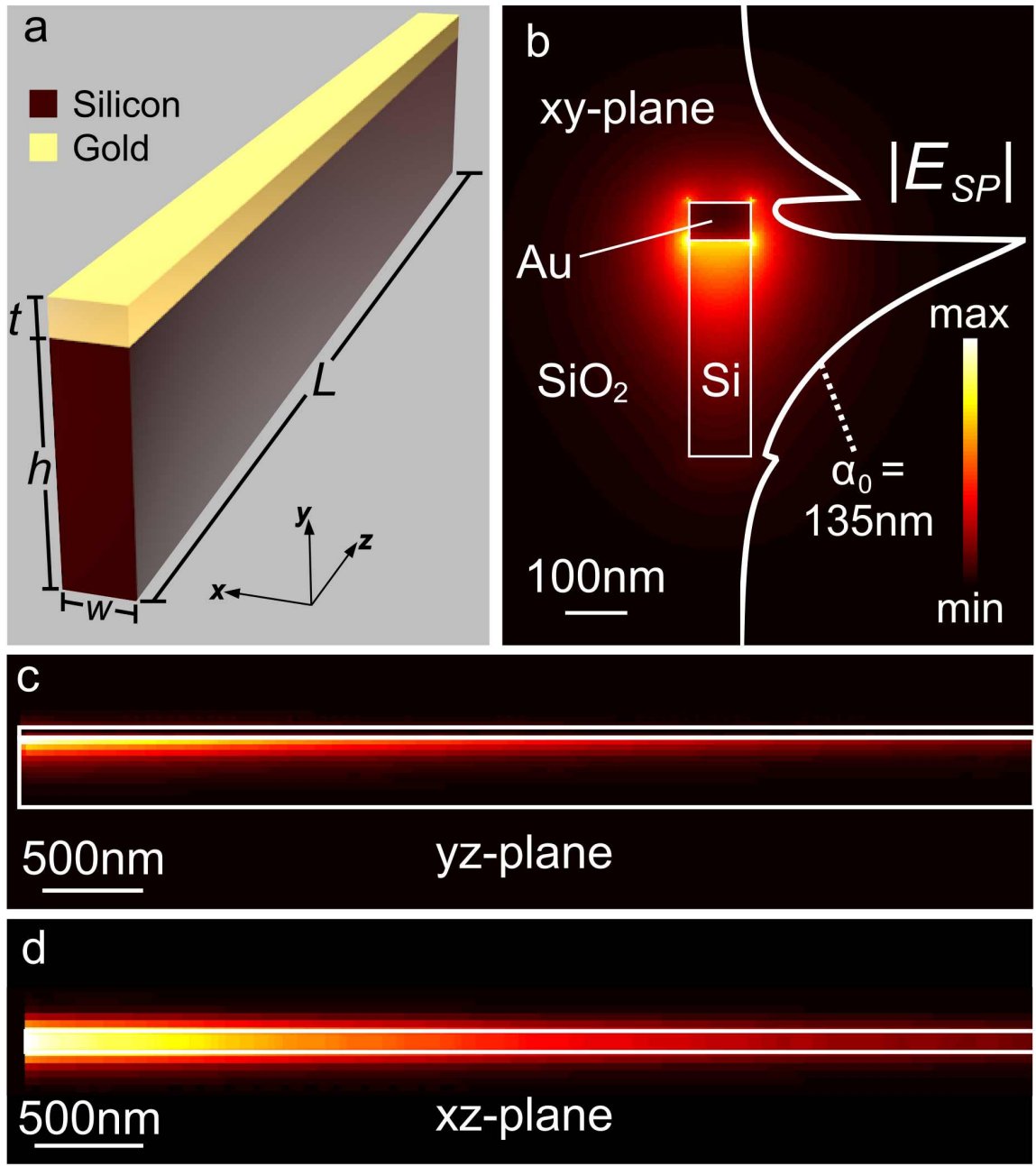

**Figure 1 | Simulation of the nanoplasmonic waveguide. a**, Schematic depiction of the Si-loaded nanoplasmonic waveguide. **b**, Time-averaged electric field amplitude distribution in the *xy*-plane of the structure. **c**, Time-averaged electric field intensity distribution in the *yz*-plane of the structure. **d**, Time-averaged electric field intensity distribution in the *xz*-plane of the waveguide at $y = h/2 = 170$nm.

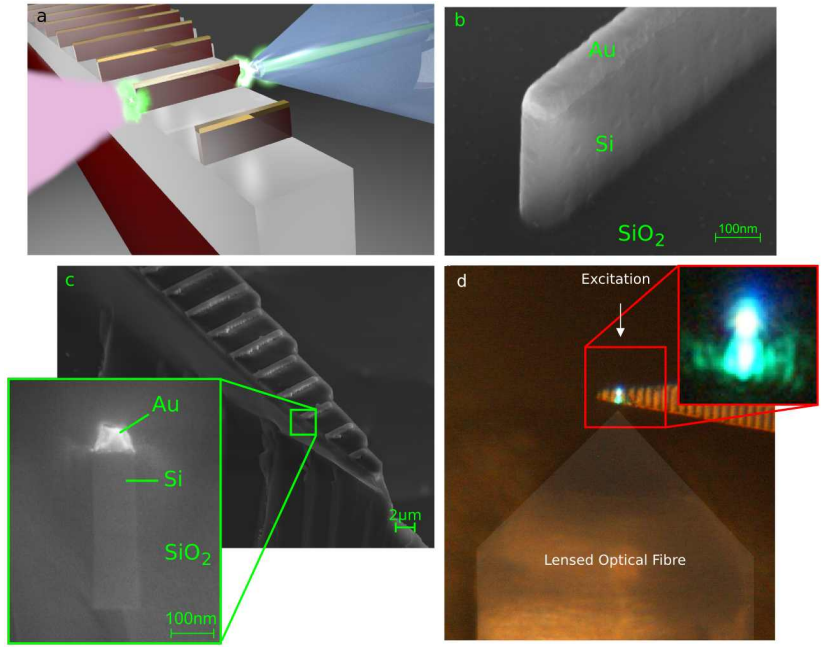

**Figure 2 | Fabricated nanoplasmonic waveguide and visible light emission. a**, Schematic depiction of the conceived device, excitation scheme, and observed visible emission. **b**, Scanning electron micrograph of a single nanoplasmonic waveguide before deposition of a $SiO_2$ cladding; **c**, Scanning electron micrograph of several nanoplasmonic waveguides fabricated on a narrow beam. A close-up of the cross-sectional geometry of the waveguide end-facet is shown in the inset. **d**, Optical microscope image of visible emission from above the plasmonic waveguides.

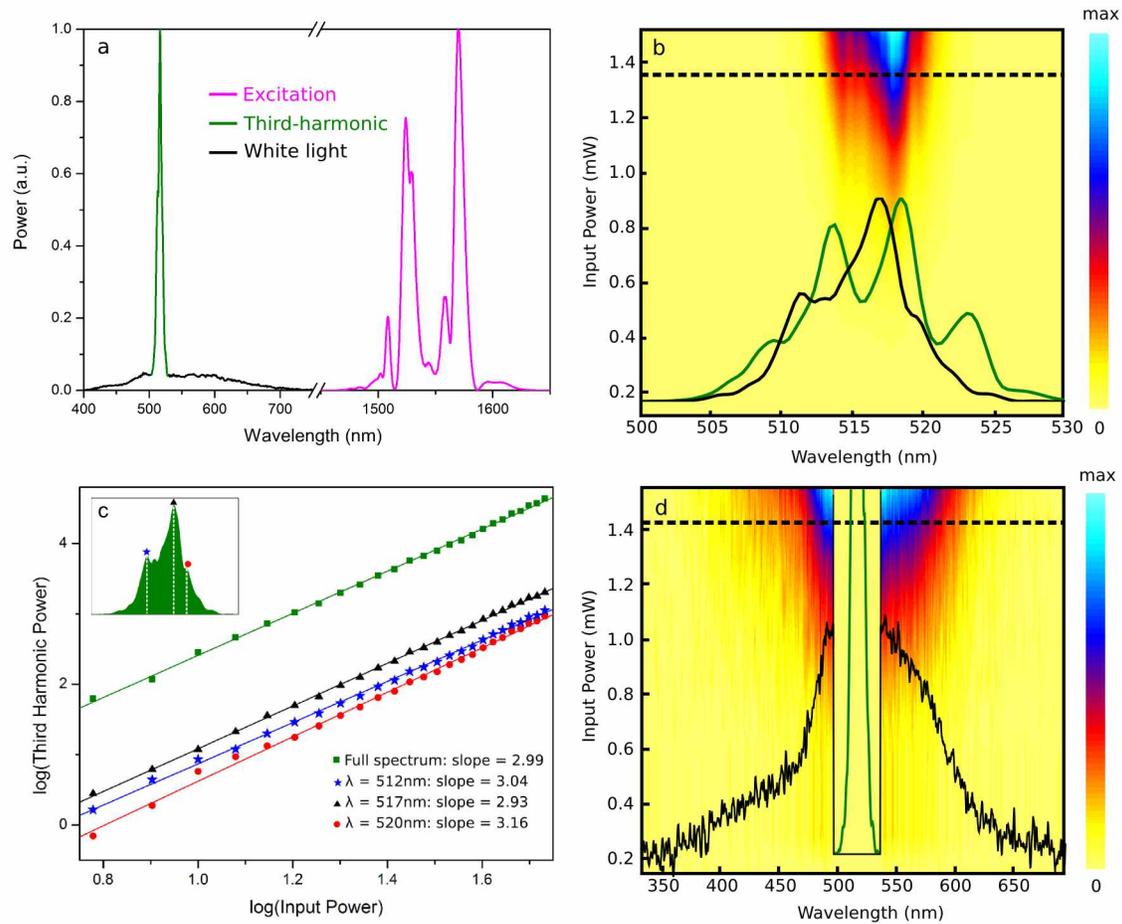

**Figure 3 | Spectral characterization of visible light emission. a**, Spectra of the ultrafast laser (magenta), third harmonic (green), and white light emission (black). **b**, Surface plot of third harmonic power as a function of wavelength and input power measured at the output facet of the nanoplasmonic waveguide. The black overlaid spectrum is a TH spectrum measured at an average input power, $P_{in}^{\omega}$ = 1.36mW, while the green spectrum is the TH spectrum calculated from the laser pulse spectrum. **c**, Log-log plots of detected third-harmonic power versus input power for the entire spectrum (green) and individual wavelengths, λ = {512, 517, 520} nm. **d**, Surface plot of white light photon counts measured at the output facet of the nanoplasmonic waveguide as a function of wavelength and input power. The black overlaid spectrum is measured at an average input power, $P_{in}^{\omega}$, = 1.42mW, and the green curves represent the position of the TH components, which have been removed for clarity.

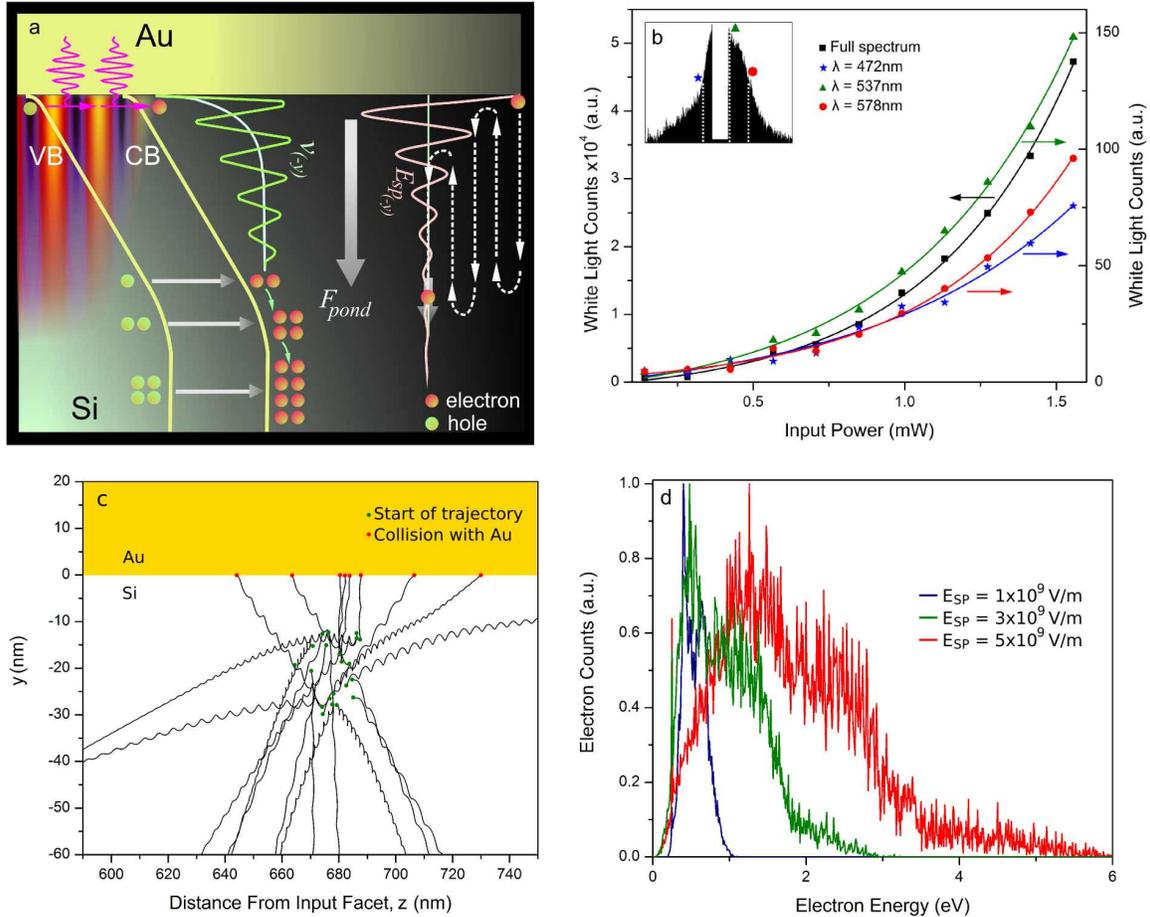

**Figure 4 | Ponderomotive electron acceleration and white light scaling. a**, Schematic depiction of physical processes relevant to the electron acceleration, including intense, asymmetric $E_{SP}$ field propagation, two-photon absorption, cycle-by-cycle acceleration of the electron in the $E_{SP}$ field leading to a net ponderomotive force, and electron and hole avalanche multiplication. **b**, Plots of experimentally measured white light counts versus input power for the entire spectrum (black) and individual wavelengths, $\lambda = \{472, 537, 578\}$ nm. **c**, Sample calculated trajectories of 21 electrons excited via two-photon absorption at random points in time, with random positions and initial velocities. The green dots represent the starting point of the electron trajectory and the red dots denote the location of an electron colliding with the gold surface. **d**, Calculated electron energy spectra for three different $E_{SP}$ values in the same range as those used in the experiments. The electron energy represents the average energy of the electrons before undergoing their first collision.

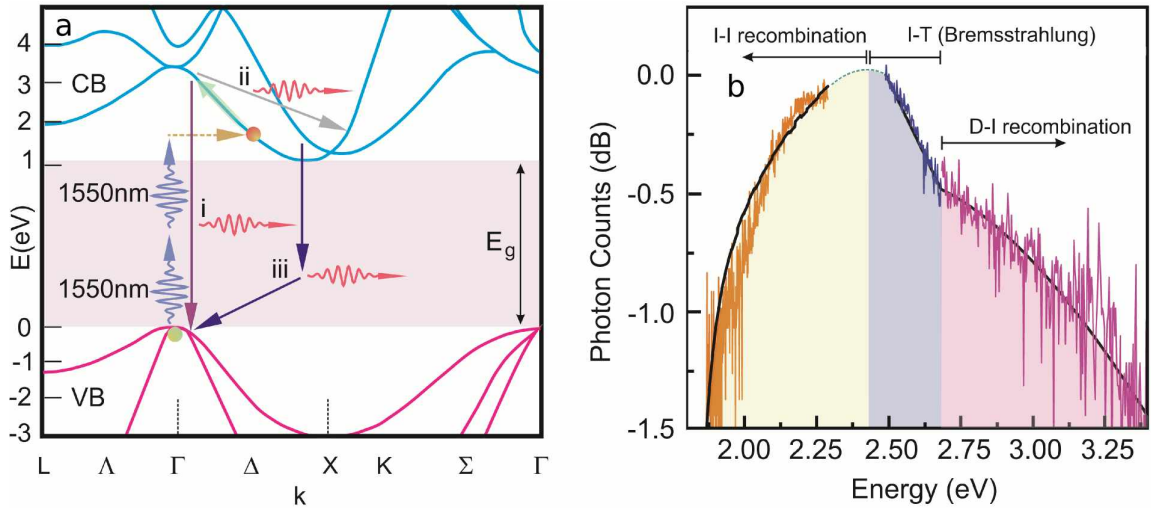

**Figure 5 | Multimechanism light emission model. a**, Schematic depiction of photon emission processes: (i) direct interband (D-I) recombination, (ii) intraband transitions (I-T) or Bremsstrahlung radiation, and (iii) indirect interband (I-I) recombination. **b**, Plot of photon counts versus photon energy, as measured in the experiments, along with the fits curves (black) obtained using a multimechanism model including D-I (magenta), I-T (blue), and I-I (orange).

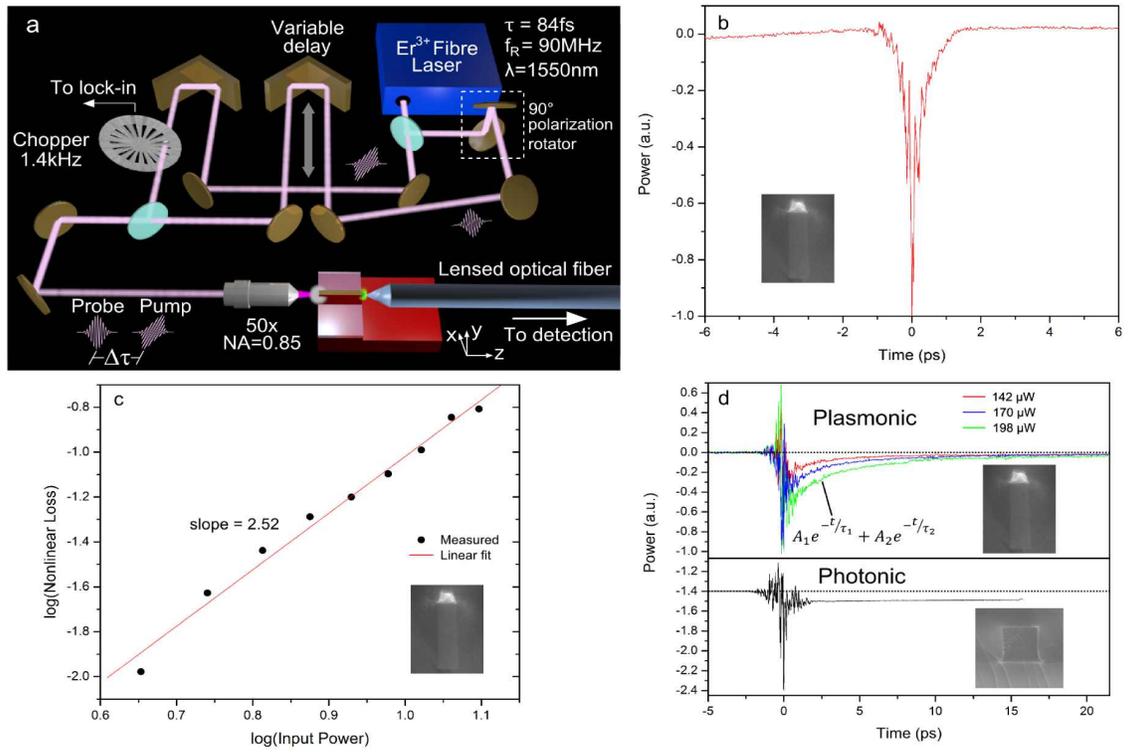

**Figure 6 | Ultrafast nonlinear characterization. a**, Artistic rendering of the ultrafast pump-probe characterization setup. **b,c** Pump-probe time-dependent power transmission for parallel and perpendicular pump-probe orientation, respectively. **d**, Log-log plot of power absorbed via FCA versus input power, with a slope of 2.52.

28